# A Quantum Physical Design Flow Using ILP and Graph Drawing

Maryam Yazdani, Morteza Saheb Zamani, Mehdi Sedighi

*Computer Engineering and IT Department, Amirkabir University of Technology, Tehran, Iran*
maryam.yazdany@aut.ac.ir, szamani@aut.ac.ir, msedighi@aut.ac.ir

Abstract: Implementing large-scale quantum circuits is one of the challenges of quantum computing. One of the central challenges of accurately modeling the architecture of these circuits is to schedule a quantum application and generate the layout while taking into account the cost of communications and classical resources as well as the maximum exploitable parallelism. In this paper, we present and evaluate a design flow for arbitrary quantum circuits in ion trap technology. Our design flow consists of two parts. First, a scheduler takes a description of a circuit and finds the best order for the execution of its quantum gates using integer linear programming (ILP) regarding the classical resources (qubits) and instruction dependencies. Then a layout generator receives the schedule produced by the scheduler and generates a layout for this circuit using a graph-drawing algorithm. Our experimental results show that the proposed flow decreases the average latency of quantum circuits by about 11% for a set of attempted benchmarks and by about 9% for another set of benchmarks compared with the best in literature.

CAD: Computer Aided Design; ILP: Integer Linear Programming; QASM: Quantum Assembly, GDToolkit: Graph Drawing Toolkit [1, 2]

## 1- Introduction

Quantum computing offers the opportunity to solve certain problems thought to be intractable on a classical machine. For example, unsorted database search [3], factorization [4] and simulation of quantum mechanical systems [5] are some classically hard problems that benefit from quantum algorithms.

In addition to significant work on quantum algorithms and underlying physics, there have been a few studies proposing quantum circuit design flows. Quantum circuit design flow consists of two processes: scheduling and layout generation. Even though it might appear that the integration of the two processes into one monolithic process can potentially lead to better final layout and scheduling, the high complexity of this design flow makes the whole process very hard. As will be shown, scheduling can be viewed as an independent process from layout generation so that an optimal scheduling can be found using ILP. The next process can then take this optimal schedule with maximum parallelism and generate a layout according to this schedule.

Ion trap technology is used as the underlying technology to study our flow. A number of physical experiments have demonstrated the implementation of small-scale quantum systems based on this technology that show promise for scalable quantum information processing [6, 7, 8, 9, 10].

According to [10], almost one third of the duration of the different tasks within a typical quantum logic process consists of ion transportation and gate pulses. Since our focus is on finding an optimum gates scheduling and layout, one can claim that by doing so, we can improve approximately one third of the whole gate operation process. This can in turn lead to considerable improvements in the overall timing.

The rest of this paper is organized as follows. An introduction to the ion trap technology is presented in Section 2, followed by an overview of the prior work in Section 3. In Section 4, the details of the proposed approach and its design flow are discussed. Section 5 shows the experimental results, and Section 6 concludes the paper.

## 2- Ion Trap Technology

Successful realization of a quantum computer is a collaborative effort that requires various disciplines and expertise. Since no single group of researchers can possess all of this required expertise, each group has to focus on one aspect of the problem with a limited scope of details while abstracting away further details that can be left to other groups. This is very similar to the current approach in semiconductor industry with some people involved in the higher level design of the circuit using some abstract models of the actual chip to be fabricated, while others are heavily involved with the detail implementation of the circuit and the associated physical challenges.

A customary practice among the papers that focus on high-level problems such as scheduling and layout generation is to use an abstract model of a particular technology as the basis for their work. This approach allows consideration of a limited set of physical constraints while focusing on the problems at a higher level of abstraction. A common underlying technology in this field is ion trap [8, 9, 11, 12]. This technology has shown good potential for scalability [9, 10, 13]. In this technology, a physical qubit is an ion and a gate is a location which a trapped ion may be operated upon by a modulated laser. A qubit may be held in place at any trap region, or it may be ballistically moved between them by applying pulse sequences to discrete electrodes which line the edges of ion traps.

There is a library of macroblocks defined in [14] that is often used in scheduling and layout generation in ion trap technology. We chose this library for two reasons. First, macroblocks abstract away some of the low-level details such as variations in the technology implementations of ion traps, the type of ion species used, specific electrode sizing and geometry, and exact voltage levels necessary for trapping and movements of ions within the macroblocks. Secondly, ballistic movement along a channel requires carefully timed application of pulse sequences to electrodes in non-adjacent traps. Using basic blocks including the interface between them requires communication only between the two blocks involved.

Figure 1 shows the library defined in [14]. Each macroblock consists of a 3x3 grid of trap regions and electrodes, with ports to allow qubit movements between the macroblocks. The black squares are gate locations, which may not be performed at intersections or turns in the ion trap technology. Each of these macroblocks may be rotated in a layout.

Some key characteristics of ion trap can be summarized as follows. In this technology, there are some rectangular channels, lined with electrodes, called "Wires". Trapped-ion devices use charged electromagnetically trapped atoms as qubit carriers. Each qubit is represented by internal electronic and nuclear states of a single ion. Laser pulses of specific frequencies addressing one or more ions apply single and multi-qubit quantum gates [15]. Atomic ions are suspended above the channel regions and can be moved ballistically [16]. By synchronized application of voltages on the channel electrodes, qubits can be moved ballistically to move data around. There is some movement control circuitry to handle any qubit communication for each wire.

Each gate location can potentially perform any operation available in the ion trap technology, so the reuse of gate locations within a quantum circuit is possible. Due to the difficulty of fabricating and controlling ion traps in the third dimension, scalable ion trap systems are two dimensional [17]. Routing channels may be shared by multiple ions as long as control circuits prevent multi-ion occupancy. Consequently, the circuit model resembles a general network. In this model, it has been shown by experiments that a right angle turn takes substantially longer than a straight channel over the same distance. Therefore movement latency of ions is dependent on not only the Manhattan distance between the source and the destination of movement but also on the geometry of the path [16].

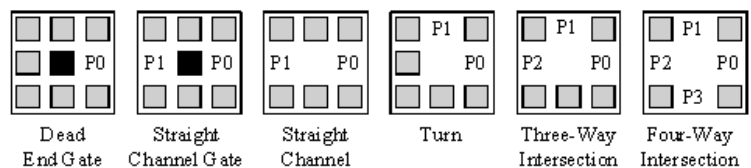

**Fig. 1** Library of basic macroblocks [14]

In Figure 1, each macroblock has a specific number of ports (shown as P0-P3) along with a set of electrodes used for ion movement and trapping. Some macroblocks contain a trap region where gates may be performed (black square) [14].

## 3- Related Work

The problem of designing a quantum circuit is based on six factors:

1. Netlist (a quantum circuit usually represented in QASM language [18])
2. Layout
3. Initial placement of qubits
4. Execution order of quantum instructions (scheduling)
5. Moving patterns (which qubit goes toward the other one to execute an instruction)
6. Routes (i.e. the paths on which every qubit moves).

According to these six factors, the problem of designing a quantum circuit can be presented in three different ways. The inputs and outputs of each kind of problems can be seen in Table 1.

On the other hand, the quantum computation community has mostly focused on a circuit model in which algorithms and architectures are tightly integrated, similar to a classical application specific integrated circuit, or ASIC design. In these approaches, the main challenge was finding a systematic method of scheduling and layout generation of quantum circuits, and there was no special focus on finding a method for generating *optimal* schedule or layout of these circuits. Consequently, some hand-optimized layouts have been used in the literature. Some prior work takes a fixed netlist and generates a layout without manipulation of the input netlist. This style causes optimization improvement to be limited because of fixing netlist after synthesis process. In some other approaches [19], local manipulations of the netlist improve circuit metrics. Metodi et al. [20] presented a Quantum Logic Array architecture which was improved in [6]. In those papers, instead of focusing on improving scheduling or layout generation, they focused on architectural aspects. Metodi et al. [21] also developed a tool to generate a schedule for physical operations given a netlist and a fixed grid-based layout (Problem 2 in Table 1). Whitney et al [14] proposed scheduling heuristics for running on grid-based and non-grid-based layouts (Problem 2) in addition to another heuristic for solving Problem 1 creating layout and scheduling according to the netlist. Mohammadzadeh et al. [19] proposed a flow for layout generation and scheduling of quantum circuits in the ion trap technology solving Problem 1. While our flow builds upon some of these ideas, we try to focus on better scheduling and layout generation algorithms.

**Table 1** Different problems for designing a quantum circuit

|  | Problem 1 | Problem 2 | Problem 3 |
|---|---|---|---|
| Problem inputs | - Netlist | - Netlist<br>- Layout<br>- Initial placement of qubits | - Netlist<br>- Layout |
| Problem outputs | - Layout<br>- Initial placement of qubits<br>- Execution order of quantum instructions<br>- Moving pattern<br>- Routes | - Execution order of quantum instructions<br>- Moving pattern<br>- Routes | - Initial placement of qubits<br>- Execution order of quantum instructions<br>- Moving pattern<br>- Routes |

# 4- Our Approach

In this section, we propose a physical design flow for quantum circuits using integer linear programming (Problem 1 in Table 1). The first step of this flow finds the best scheduling for the circuit using ILP and the second step designs the layout structure according to this scheduling. During the layout design, the initial placement of qubits, moving patterns and routes are also determined.

Our design flow is summarized in Figure 2. We start with a technology-dependent netlist, produced by a synthesis tool or specified by the designer. ILP equations are generated to schedule the netlist and an ILP solver solves the equations and finds the optimal scheduling. In the next step, this scheduling is used and a graph of qubit flow, called qubit flow graph (QFG), is constructed. Then this graph is given to a graph drawing tool, called GDToolkit [1], to obtain an orthogonal graph with minimum number of turns in edges and minimum length for edges. This orthogonal graph is used as the basic pattern of the layout. The edges and nodes of this graph are converted to the macroblock introduced in Section 2 to generate an optimized layout with respect to the number of bends and short routes. Next, an initial placement of qubits, moving patterns and the routes are determined.

In the following subsections, scheduling and layout generation steps are described in details.

The following lemma [22] and definition are necessary to describe our flow.

***Lemma:*** Let gates *A* and *B* be controlled gates such as controlled-NOT, controlled-Z etc. , gate *A* have control set $C_A$[1] and target $T_A$ and gate *B* have control set $C_B$ and target $T_B$. These two gates are *exchangeable* if, and only if, one of the following conditions is satisfied [22]

1. $T_A \notin C_B$ and $T_B \notin C_A$                if A and B are of the same type

2. $T_A \notin C_B$ and $T_B \notin C_A$ and $T_B \notin T_A$     if A and B are of different types

For example, in Figure 3b, gates 6 and 7 can be exchanged.

***Definition:*** We call two gates *dependent* or have *data dependency* if one of them cannot be executed before the other one. This means that they are not *exchangeable*. For example, in Figure 3b, gate 10 is dependent on gate 3.

## 4-1- Scheduling

We start with a technology-dependent netlist. Figure 3a shows the QASM instruction sequence of the circuit [[9,3,2]] (introduced in [23]) consisting of Hadamard gates (H) and controlled bit-flips (CX) operating on qubits q0 to q8 with each instruction labeled by a number. Figure 3b shows the equivalent sequence of operations in the form of schematic quantum circuit format. Either of these may be translated into the dataflow graph shown in Figure 3c where each node represents a QASM instruction (as labeled in Figure 3a) and each arc represents a qubit dependency. According to the lemma presented in Section 4, instructions 4 and 5, 6 and 7, 7 and 8, 5 and 9 and some other instructions have common qubits, and therefore, they are in some way dependent on each other and cannot be executed simultaneously as in the ion trap technology, an ion can take part in only one operation at a time. However, since they are exchangeable, they may perform in any order in the optimal schedule. Apart from this, there are some real dependencies as in instructions 16 and 19 that are dependent on instruction 4 and cannot be executed before instruction 4 but instruction 16 may be executed before or after 19. A Modified version of dataflow graph (introduced in [14]) is used here in which a directed edge from $i$ to $j$ indicates that instruction $j$ is dependent on $i$ and cannot be executed before it. While drawing the dataflow graph, it is not clear in which order these instructions will be executed in the optimal scheduling and only the data dependencies can be demonstrated in this graph (Figure 3c). Using this dataflow graph, we may write the ILP equations as described in the following subsection.

---

[1] $C_A$ is an empty set in case of an uncontrolled gate

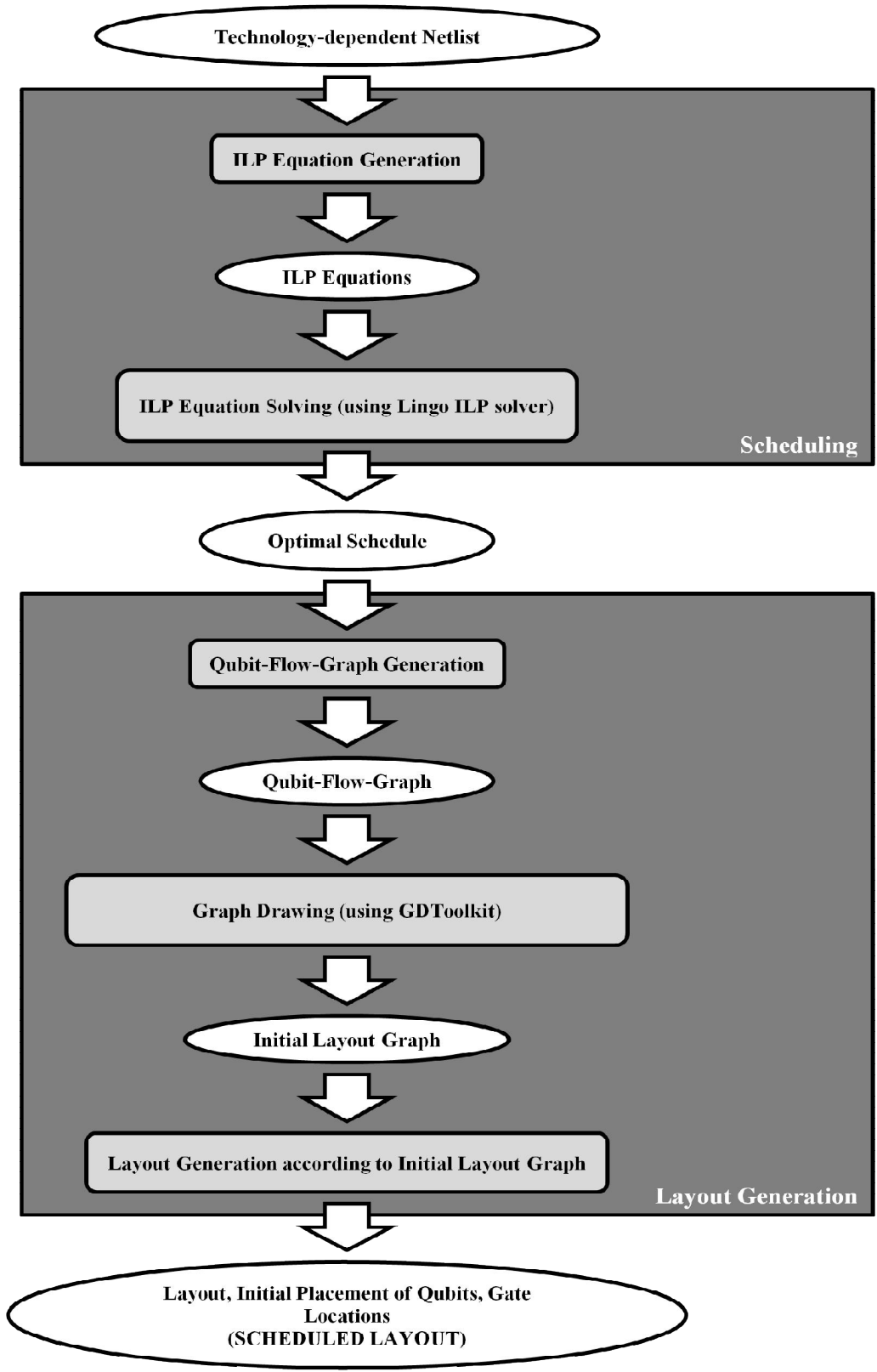

**Fig. 2** Our design flow. The gray parts do the needed process on the white parts

| (1)  | H q0     |
|------|----------|
| (2)  | H q1     |
| (3)  | H q2     |
| (4)  | CX q5,q7 |
| (5)  | CX q5,q8 |
| (6)  | CX q4,q6 |
| (7)  | CX q3,q6 |
| (8)  | CX q3,q7 |
| (9)  | CX q3,q8 |
| (10) | CX q2,q3 |
| (11) | CX q2,q4 |
| (12) | CX q2,q6 |
| (13) | CX q2,q7 |
| (14) | CX q2,q8 |
| (15) | CX q1,q3 |
| (16) | CX q1,q5 |
| (17) | CX q1,q7 |
| (18) | CX q0,q3 |
| (19) | CX q0,q5 |
| (20) | CX q0,q8 |

(a)

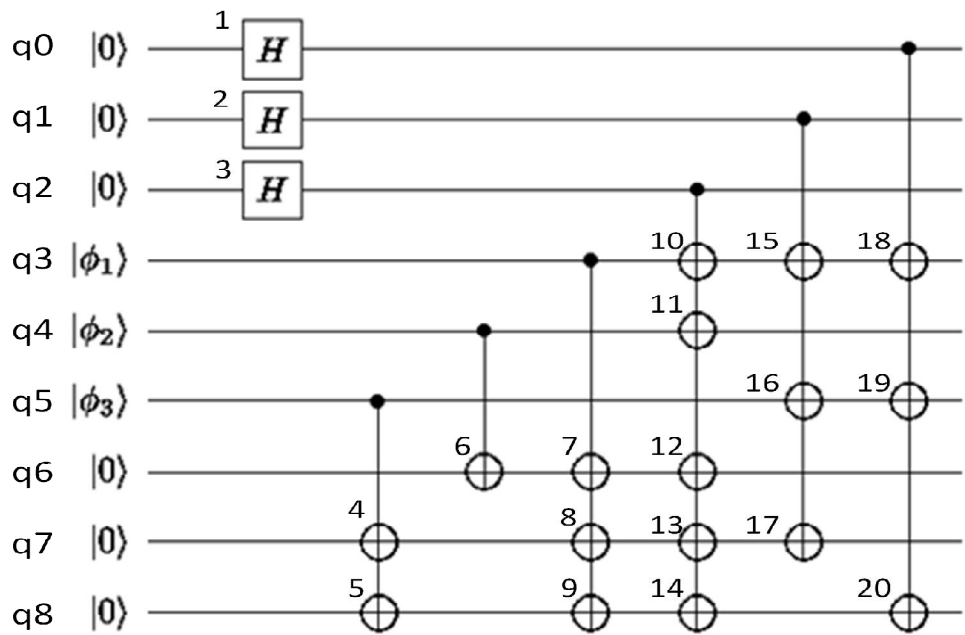

(b)

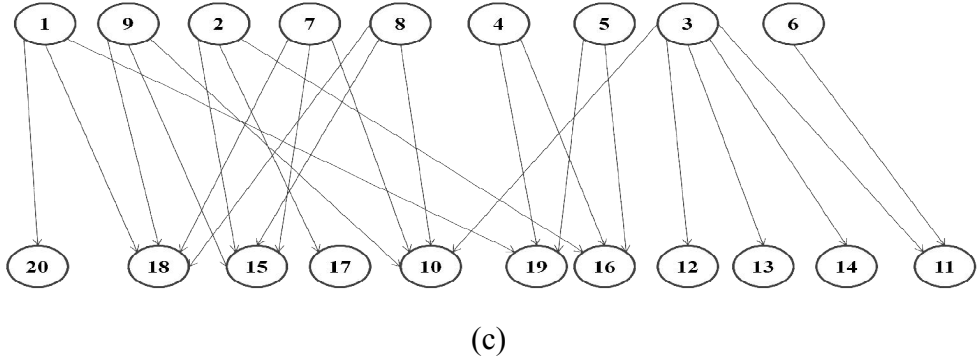

(c)

**Fig. 3** a) The QASM instruction sequence of circuit [[9,3,2]]. b) The quantum circuit equivalent to the instruction set of circuit [[9,3,2]]. c) The dataflow graph equivalent to the instruction sequence in (a). Each node represents an instruction, as labeled in (a). Each arc represents a qubit dependency

According to the above netlist, the common qubits of instructions are demonstrated in Table 2. For example, q3 is the common qubit of instructions 7, 8, 9, 10, 15 and 18. In the ion trap technology with two-qubit gates, no more than one instruction can use a qubit at the same time. Consequently, in the best case, each of the instructions 7, 8, 9, 10, 15, 18 should be done in a separate step. In this netlist, at most six instructions use the same qubit. This means that this netlist cannot be scheduled in less than 6 steps; consequently there are 6 scheduling steps in the dataflow graph. According to this data flow graph, the instructions are scheduled using ASAP[2] and ALAP[3] algorithms and then the slack of each instruction is obtained. The earliest/latest time at which an instruction can be executed, determined by ASAP/ALAP is called $t_{ASAP}/t_{ALAP}$. We use these for writing the ILP equations.

**Table 2** Common qubits of instruction

| Common qubit | Instructions |
|---|---|
| q0 | 1, 9, 18, 19, 20 |
| q1 | 2, 15, 16, 17 |
| q2 | 3, 10, 11, 12, 13, 14 |
| q3 | 7, 8, 9, 10, 15, 18 |
| q4 | 6, 11 |
| q5 | 4, 5, 16, 18 |
| q6 | 6, 7, 12 |
| q7 | 4, 8, 13, 17 |
| q8 | 5, 9, 14, 20 |

---

[2] As Soon As Possible
[3] As Late As Possible

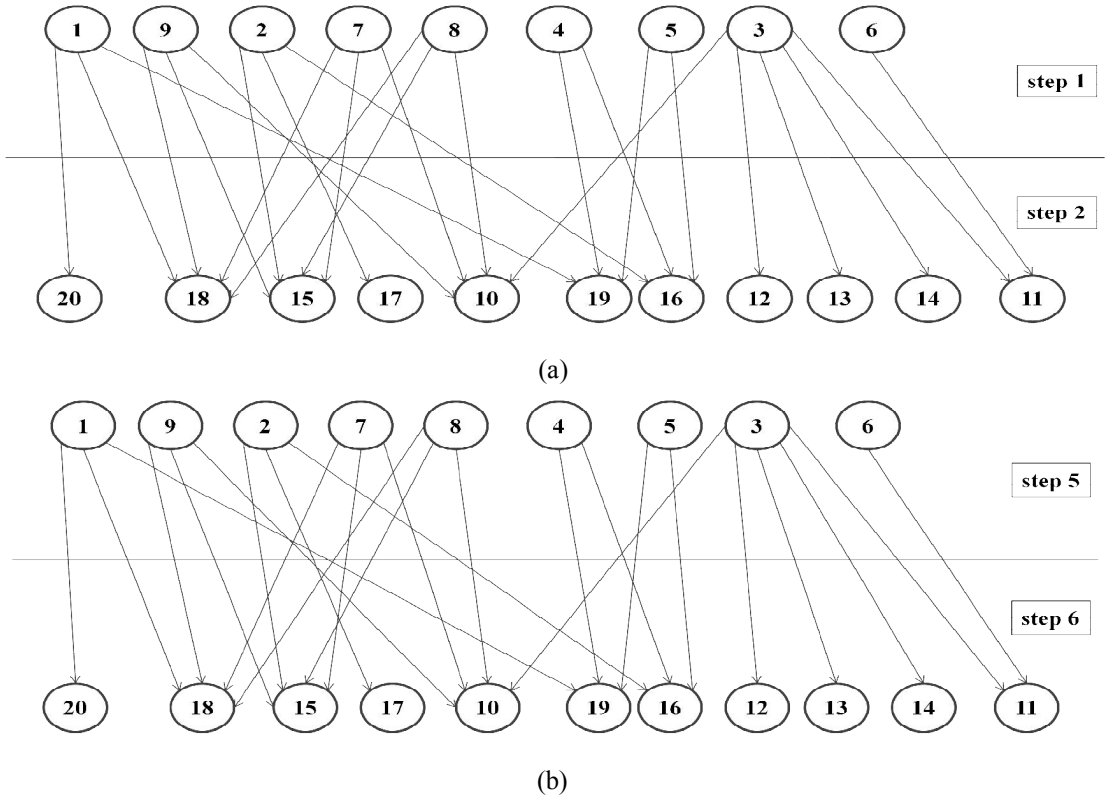

**Fig. 4** a) Dataflow graph of circuit [[9,3,2]] scheduled by ASAP. b) Dataflow graph of circuit [[9,3,2]] scheduled by ALAP

*4-1-1- Equation series 1*

The first series of ILP equations ensures that each instruction is executed only once.

$$\forall i \sum_{l=t_{ASAP}}^{t_{ALAP}} x_i(l) = 1 \qquad \text{Equation series (1)}$$

where $x_i(l)$ shows that instruction $i$ is executed in step $l$.

For example, for instruction number 6 in Figure 4, $t_{ASAP}$ is 1, and $t_{ALAP}$ is 5. Then the equation is as follows:

$$x_6(1)+x_6(2)+x_6(3)+x_6(4)+x_6(5)=1$$

This means that instruction number 6 can be executed no sooner than stage 1 and no later than stage 5, and it can only be executed in one and only one of these five stages.

### 4-1-2- Equation series 2

The second series of equations considers resource constraints. Here, the resources are qubits and these equations guarantee that at one moment, two different instructions are not applied on the same qubit. To this end, Equation series (2) are written in such a way that for each pair of instructions $i$ and $j$ sharing at least a qubit, if they can potentially be scheduled in the same stage $l$, i.e. $t_{ASAP} \leq l \leq t_{ALAP}$, they should be mutually exclusive for that stage:

$$x_i(l) + x_j(l) \leq 1 \qquad \text{Equation series (2)}$$

### 4-1-3- Equation series 3

The third series of equation is generated according to the data flow graph. We should write Equation series (3) for each pair of instructions which have data dependency (defined in Section 4). These equations ensure that if instruction $i$ is dependent on instruction $j$, then $i$ cannot execute before $j$.

$$\forall i,j; if\,(i\ depends\ on\ j)\,then$$
$$\sum_{l=t_{ASAP}}^{t_{ALAP}} l \times x_j(l) < \sum_{l=t_{ASAP}}^{t_{ALAP}} l \times x_i(l) \qquad \text{Equation series (3)}$$

An ILP solver can take these series of equations and find the optimal solution in the form of $x_i(l) = 1$ which means instruction $i$ should be executed in stage $l$, or $x_i(l) = 0$ which means instruction $i$ should not be executed in stage $l$.

For example, the result of ILP-based scheduling for circuit [[9,3,2]] (Figure 3) is shown in Table 3.

Table 3 Result of the ILP-based scheduling for circuit [[9,3,2]]

| Stages of scheduling | Instructions which should be executed in this stage |
|---|---|
| 1 | 1,2,3,4,9 |
| 2 | 5,7,13 |
| 3 | 6,8,14,16 |
| 4 | 10,19 |
| 5 | 12,15,20 |
| 6 | 11,17,18 |

## 4-2- Layout generation

In the layout generation step, the result of scheduling is used and a graph is generated accordingly which shows the flow of qubits in the circuit. In this graph, each node represents an instruction and an edge between node *i* and node *j* shows that a qubit moves from instruction *i* to instruction *j*. The instructions which should be executed at the same time are assigned to the same stage in this graph based on the results of the scheduler (Table 3). The qubit flow graph for circuit [[9,3,2]] is shown in Figure 5.

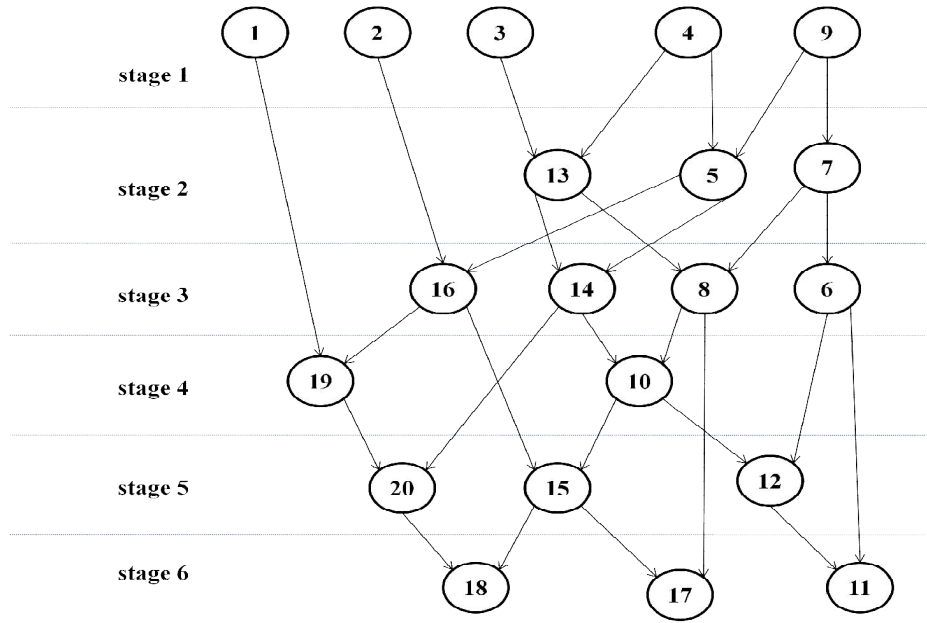

**Fig. 5** Qubit flow graph for circuit [[9,3,2]]

This graph is given to a graph drawing algorithm, called Giotto algorithm [1] which has been implemented in GDToolkit. In the ion trap technology, the delay of turns is modeled as 10 times the delay of straight paths [24]. This tool generates an orthogonal graph with minimum turns and minimum length of edges to minimize the delay. The output of GDToolkit for circuit [[9,3,2]] is shown in Figure 6.

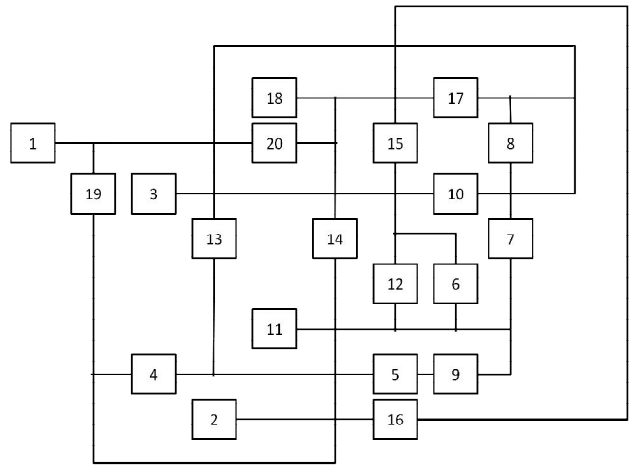

**Fig. 6** The orthogonal graph generated by GIOTTO algorithm

Here, nodes and edges are shown by squares and rectilinear lines, respectively.

Then this graph is converted to a layout by using basic macroblocks according to Table 4.

**Table 4** Equivalent macroblocks and graph components to convert an orthogonal graph to a layout

| Macroblock | Node or edge in the graph |
|---|---|
| 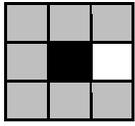 | 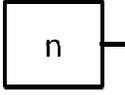 |
| 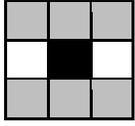 | 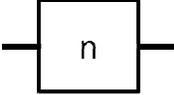 |
| 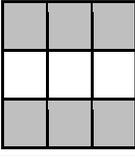 | 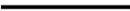 |
| 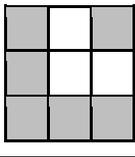 | 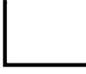 |
| 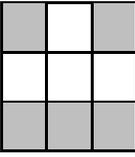 | 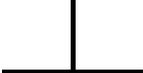 |
| 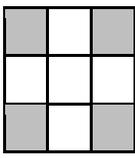 | 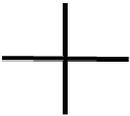 |

Layout of circuit [[9,3,2]] generated by our approach is shown in Figure 7. In this layout, the grey squares are the electrodes, the white squares are the empty spaces between the electrodes in which the qubits can be trapped or moved, the black squares are the gate locations and the numbers show the gate numbers in their specified locations.

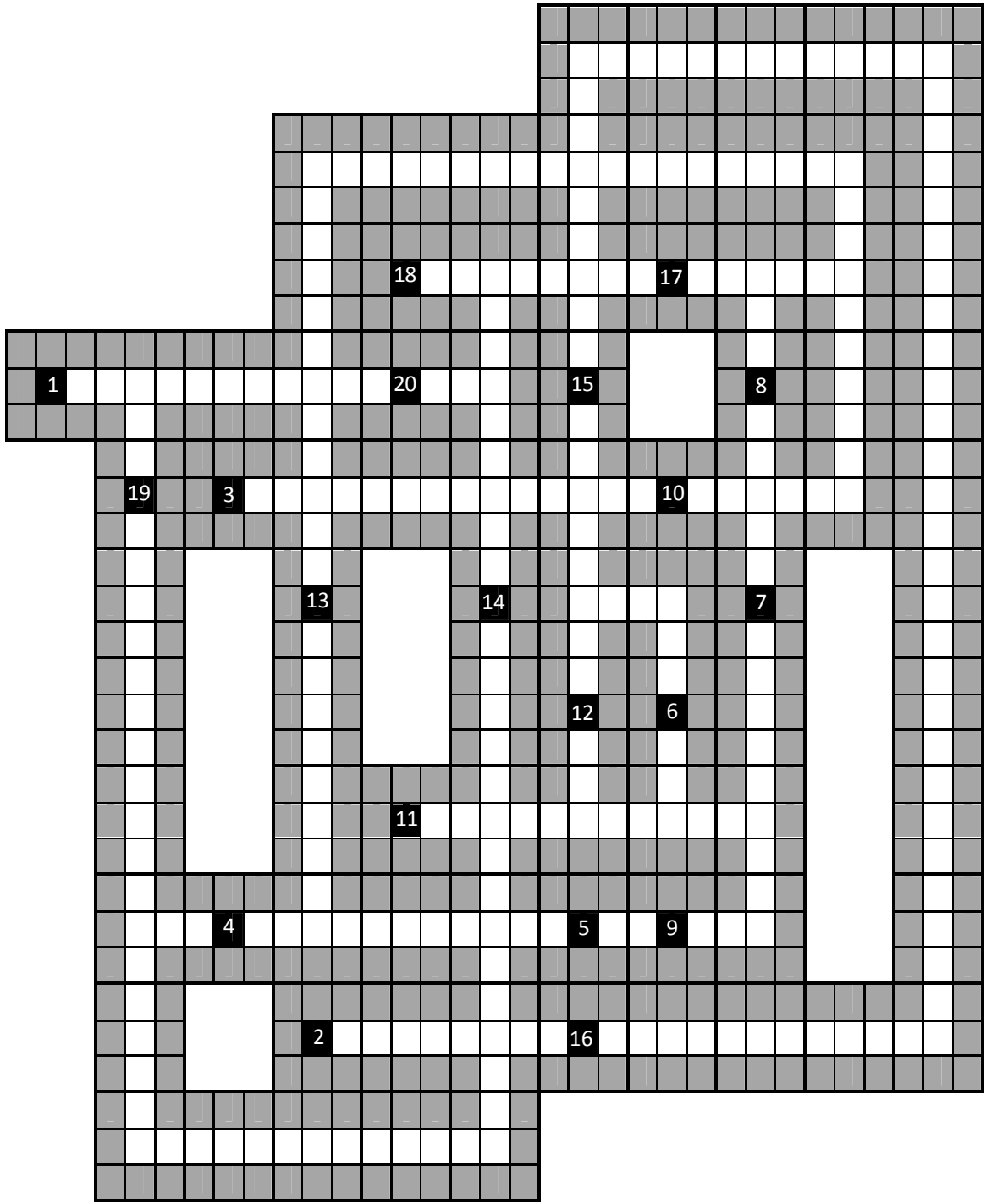

**Fig. 7** Layout and gate locations of circuit [[9,3,2]] created by our approach

The initial placement of each qubit is chosen to be the location of the gate which uses that qubit for the first time. The movement pattern and the routes are also determined according to the graph in Figures 5 and 6.

# 5- Experimental Results

To compare our results with those produced by the best in the literature, we used the benchmarks the results of which have been reported in [14], [26] and [27]. Details of the benchmarks and the latency values for various physical operations in the ion trap technology can be seen in Tables 5 and 6. The benchmarks used in [14] and [26] were chosen from [23] and those used in [14] and [27] were taken from [24]. Therefore, the results are compared with them in Table 7 and Table 8 for the two sets of benchmarks. The gate library consists of Hadamard, CNOT (Controlled-X), Controlled-Y, Controlled-Z and $C^n$NOT where $C^n$NOTs are decomposed into CNOTs, Controlled-Vs and Controlled-V†s. So, the circuits are composed of H, CNOT, $C^n$NOT, Controlled-Y and Controlled-Z gates. $C^n$NOT gates (including Toffoli gates) are decomposed according to [28, 29]. Then we schedule the circuit and generate the layout.

We also used another gate library motivated and supported by fault tolerance protocols. To achieve this, we used the decomposition of Toffoli gates into T, $T^\dagger$, S, H and CNOTs according to [28].

C++ is used as the programming language for generating the ILP equations as the input to the ILP solver. Lingo 13 is used as the ILP solver. Time and memory used are according to the reports of Lingo 13 [30]. For the layout generation phase, GDToolkit is used as the graph drawing tool.

All results of this section are obtained on a 2.2 GHz Intel Core 2 Duo CPU with 2 gigabytes of memory.

**Table 5** Benchmarks used for evaluation of our approach

| # | Circuit name [23, 25] | Qubit count | Gate count |
|---|---|---|---|
| 1 | [[5,1,3]] L1 encode | 5 | 13 |
| 2 | [[7,1,3]] L2 encode | 7 | 18 |
| 3 | [[9,1,3]] L1 encode | 9 | 28 |
| 4 | [[10,3,3]] L1 encode | 10 | 44 |
| 5 | [[11,1,5]] L1 encode | 11 | 47 |
| 6 | [[13,1,5]] L1 encode | 13 | 64 |
| 7 | [[16,3,5]] L1 encode | 16 | 89 |
| 8 | [[18,1,7]] L1 encode | 18 | 102 |
| 9 | [[21,1,7]] L1 encode | 21 | 140 |
| 10 | [[24,3,7]] L1 encode | 24 | 205 |
| 11 | 2 of 5-D2 | 7 | 19 |
| 12 | 2 of 5-D3 | 6 | 29 |
| 13 | ham 7-D2 | 7 | 30 |
| 14 | ham 7-D1 | 7 | 34 |
| 15 | hwb 6-D3 | 6 | 63 |
| 16 | ham15-D3 | 15 | 126 |

**Table 6** The latency values for various physical operations in the ion trap technology [24]

| Physical operation | Latency ($\mu s$) |
|---|---|
| One-qubit gate | 1 |
| Two-qubit gate | 10 |
| Measurement | 50 |
| Zero prepare | 51 |
| Straight move | 1 |
| turn | 10 |

The latency of the circuits resulted from our approach, compared with those produced by [14] and [26] are shown in the Table 7 and those results compared with those produced by [14] and [27] are shown in the Table 8.

**Table 7** The latency of the benchmark circuits achieved by the proposed flow compared with the best in the literature

| # | Circuit name [23] | Latency ($\mu s$) | | | Improvement compared with [14] (%) | Improvement compared with [26] (%) | ILP Solver runtime (s) | Memory used by ILP solver (KB) |
|---|---|---|---|---|---|---|---|---|
| | | Prior physical design flow [14] | Prior physical design flow [26] | Proposed physical design flow | | | | |
| 1 | [[5,1,3]] L1 encode | 320 | 294 | 239 | 25.31 | 18.7 | <1 | 68 |
| 2 | [[7,1,3]] L1 encode | 331 | 312 | 263 | 20.54 | 15.7 | <1 | 87 |
| 3 | [[9,1,3]] L1 encode | 574 | 405 | 383 | 33.27 | 5.43 | <1 | 185 |
| 4 | [[10,3,3]] L1 encode | 960 | 800 | 738 | 23.12 | 7.75 | 5 | 810 |
| 5 | [[11,1,5]] L1 encode | 842 | 728 | 659 | 21.73 | 9.47 | 1 | 532 |
| 6 | [[13,1,5]] L1 encode | 1281 | 1085 | 1012 | 20.96 | 6.72 | 8 | 965 |
| 7 | [[16,3,5]] L1 encode | 1757 | 1571 | 1383 | 21.28 | 11.96 | 57 | 1037 |
| 8 | [[18,1,7]] L1 encode | 1612 | 1417 | 1237 | 23.26 | 12.7 | 36 | 924 |
| 9 | [[21,1,7]] L1 encode | 3068 | 2245 | 2006 | 34.61 | 10.64 | 127 | 1850 |
| 10 | [[24,3,7]] L1 encode | 4587 | 4058 | 3670 | 19.99 | 9.56 | 356 | 1968 |
| | average | | | | 24.4 | 10.86 | | |

**Table 8** The latency of the benchmark circuits achieved by the proposed flow compared with the best in the literature

| # | Circuit name [25] | Latency ($\mu s$) | | | Improvement compared with [14] (%) | Improvement compared with [27] (%) | ILP solver run time (s) | Memory used by ILP solver (k) |
|---|---|---|---|---|---|---|---|---|
| | | Prior physical design flow [14] | Prior physical design flow [27] | Proposed physical design flow | | | | |
| 1 | 2 of 5-D2 | 3940 | 3881 | 3525 | 10.53 | 9.17 | <1 | 172 |
| 2 | 2 of 5-D3 | | | 3489 | 11.44 | 10.10 | 4 | 921 |
| 3 | ham7-D2 | 2926 | 2763 | 2549 | 12.88 | 7.74 | 25 | 667 |
| 4 | ham7-D1 | 4005 | 3896 | 3558 | 11.16 | 8.67 | 2 | 592 |
| 5 | hwb6-D3 | 6612 | 6436 | 5958 | 9.89 | 7.42 | 9092 | 3644 |
| 6 | ham15-D3 | 9536 | 9227 | 8294 | 13.01 | 10.11 | 263 | 1930 |
| | Average | | | | 11.48 | 8.86 | | |

The 5[th] and 6[th] columns in these tables show the latency improvement resulted from our approach compared with the mentioned references. ILP solver runtime and its memory usage for each circuit are shown in the last two columns, respectively.

As can be seen, an improvement of more than 11% (on average) has been achieved for the latency of the first set of benchmarks compared with the best in the literature (Table 7) and the latency of benchmarks have been improved by about 9% (on average) compared with the best in the literature (Table 8).

The latency of the circuits resulted from our approach, using the decomposition into the gate library motivated by the fault tolerance protocols are shown in Table 9.

**Table 9** The latency of the benchmark circuits achieved by the proposed flow using the gate library motivated by fault tolerance protocols

| # | Circuit name [25] | Number of gates | | | | | | Latency ($\mu s$) (Proposed design flow) | ILP solver run time (s) | Memory used by ILP solver (k) |
|---|---|---|---|---|---|---|---|---|---|---|
| | | H | T | T$^\dagger$ | S | CNOT | Total number of elementary gates | | | |
| 1 | 2 of 5-D2 | 14 | 21 | 28 | 7 | 47 | 117 | 3570 | <1 | 634 |
| 2 | 2 of 5-D3 | 40 | 60 | 80 | 20 | 129 | 329 | 5283 | <1 | 2187 |
| 3 | ham7-D2 | 26 | 39 | 52 | 13 | 92 | 222 | 4133 | 7.5 | 1178 |
| 4 | ham7-D1 | 38 | 57 | 76 | 19 | 130 | 320 | 6304 | 42 | 2486 |
| 5 | hwb6-D3 | 58 | 87 | 116 | 29 | 199 | 489 | 10461 | 238 | 2974 |
| 6 | ham15-D3 | 60 | 90 | 120 | 30 | 278 | 578 | 12427 | 211 | 6967 |

It should be noted that the decomposition of a Toffoli gate into H, T, T$^\dagger$, S and CNOT gates, results in more elementary gates than the decomposition into Controled-V, Controled-V$^\dagger$ and CNOT gates, therefore, the latency values in Table 9 are not comparable with those in Table 8.

The ILP-based optimization finds the exact optimal result in the whole search space. However, the search space increases almost exponentially as the size of the input netlist grows and so does the run time of the ILP solver. This results in the restriction of our method to smaller circuits; for larger benchmarks, techniques like hierarchical partitioning can be developed based on this approach.

For the attempted benchmarks, the time consumed by the rest of the algorithm (i.e., of the ILP equations generation and GDToolkit) is remarkably less than the time consumed by the ILP solver, and therefore, only the run time of the ILP solver is reported in Tables 7, 8 and 9.

In the literature, a constructive algorithm for synthesizing linear depth stabilizer circuits in the LNN architecture has been presented[31]. The result of the proposed design flow for [[23_1_7]] ([21] Fig. 8b) is 108 cycles, which is better than [21] (177 cycles) but not [31] (68 cycles), and this is because, the proposed design flow does not interfere with the synthesis process. The primary focus of this paper is on the scheduling and layout generation and not on the synthesis. As such, the proposed design flow receives the circuit netlist and determines the temporal order of gate executions i.e. scheduling and does not manipulate the netlist (as [31] does). Another important difference between the proposed design flow and [31] is that the latter is devoted to the synthesis of stabilizer circuits in the LNN architecture. This is a very limited scope compared to the proposed approach that can be applied to any arbitrary quantum circuit.

In order to try more experiments, we applied our approach to Cat state generation circuits. In general, the following represents Cat state:

$$|CAT\rangle = |\Psi\rangle = \frac{(|00\ldots0\rangle + |11\ldots1\rangle)}{\sqrt{2}}$$

Cat state generation circuit for 7 qubits [32] and the netlist of the circuit can be seen in Figure 8.

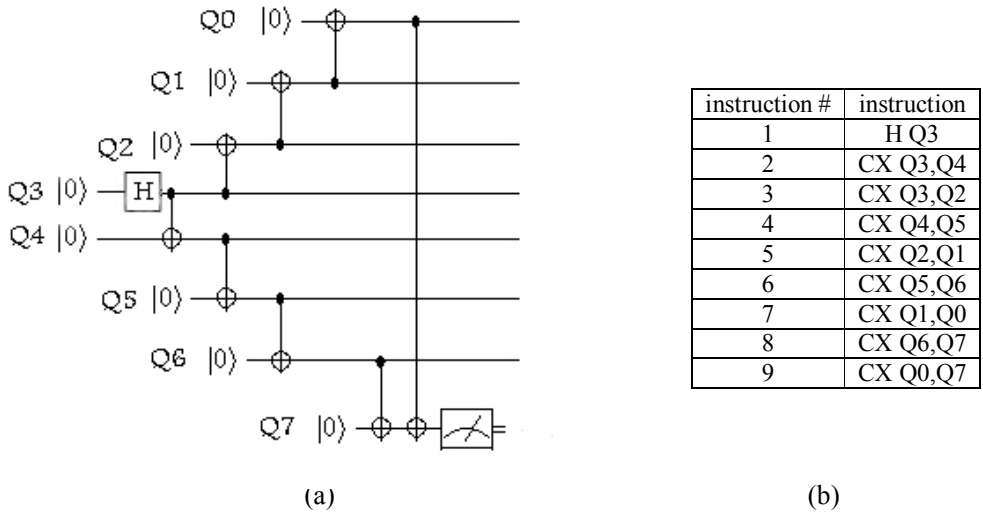

(a)                 (b)

**Fig. 8** a) The quantum circuit of 7-qubit Cat state generation. b) The QASM instruction sequence for the circuit.

The results of scheduling and the QFG for this circuit are shown in Table 10 and Figure 9.

**Table 10** Result of the ILP-based scheduling for 7-qubit Cat state generation circuit

| Stages of scheduling | Instructions which should be executed in this stage |
|---|---|
| 1 | 1 |
| 2 | 2 |
| 3 | 3,4 |
| 4 | 5,6 |
| 5 | 7,8 |
| 6 | 9 |

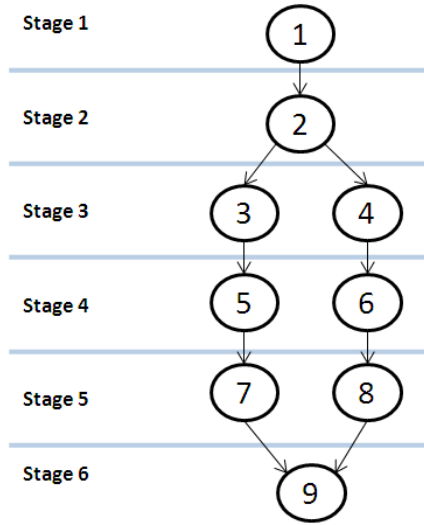

**Fig. 9** Qubit flow graph for 7-qubit Cat state generation circuit

The output of GDToolkit for the 7-qubit Cat state generation circuit is shown in Figure 10.

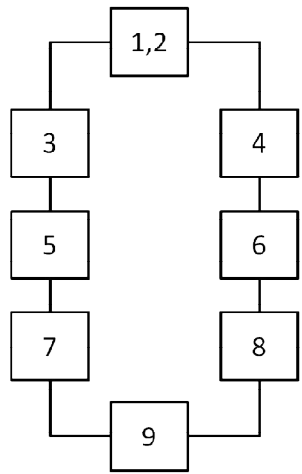

**Fig. 10** The orthogonal graph generated by GIOTTO algorithm

The layout of the 7-qubit Cat state generation circuit produced by our approach is shown in Figure 11.

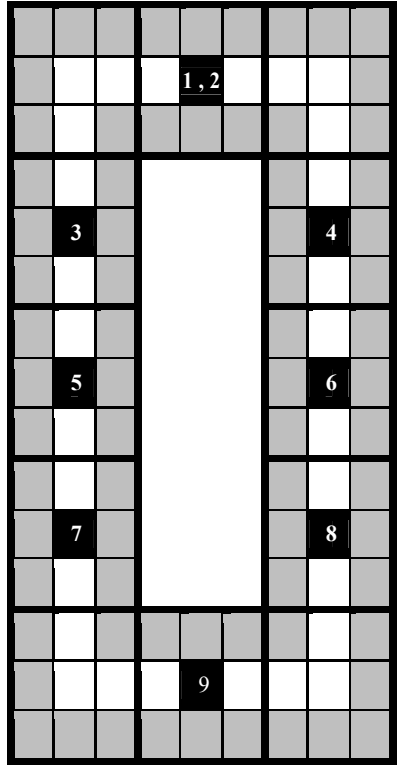

**Fig. 11** Layout and gate locations for the 7-qubit Cat state generation circuit produced by our approach

The initial placement of the qubits is also shown in Figure 12. As shown in this figure, gate #1 is executed on Q3, then Q4 moves towards Q3 and gate #2 is executed. After that, Q4 moves towards Q5 and Q2 moves towards Q2, and this continues until Q0 and Q7 move towards each other and gate #9 is executed on them in the gate location at the bottom of the layout.

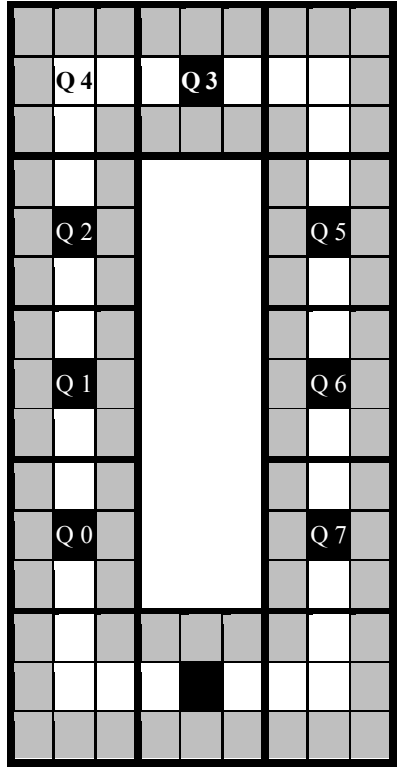

**Fig. 12** Initial placement of qubits for 7-qubit Cat state generation circuit

As another example, Cat state generation circuit for 4 qubits [32] and the netlist of the circuit are shown in Figure 13.

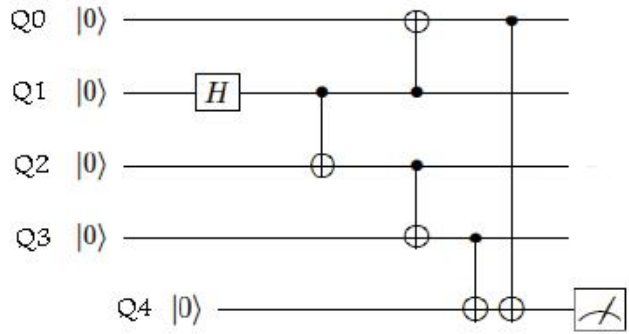

| instruction # | instruction |
|---|---|
| 1 | H Q1 |
| 2 | CX Q1,Q2 |
| 3 | CX Q1,Q0 |
| 4 | CX Q2,Q3 |
| 5 | CX Q3,Q4 |
| 6 | CX Q0,Q4 |

(a)                  (b)

**Fig. 13** a) The quantum circuit equivalent to the instruction set of 4-qubit Cat state generation circuit. b) The QASM instruction sequence of 4-qubit Cat state generation circuit.

The result of scheduling and QFG for this circuit is shown in Table 11 and Figure 14.

**Table 11** Result of the ILP-based scheduling for 4-qubit Cat state generation circuit

| Stages of scheduling | Instructions which should be executed in this stage |
|---|---|
| 1 | 1 |
| 2 | 2 |
| 3 | 3,4 |
| 4 | 5 |
| 5 | 6 |

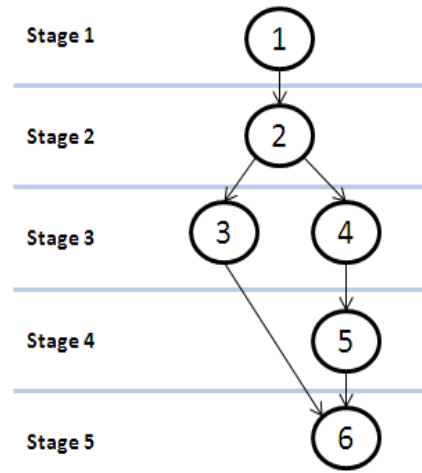

**Fig. 14** Qubit flow graph for 4-qubit Cat state generation circuit

The output of GDToolkit for 4-qubit Cat state generation circuit is shown in Figure 15.

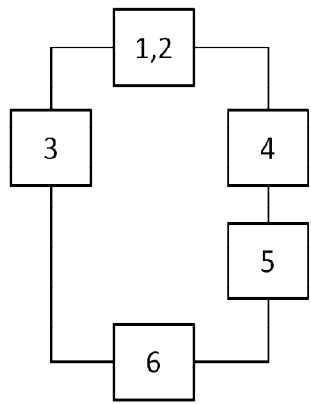

**Fig. 15** The orthogonal graph generated by GIOTTO algorithm

The layout of 4-qubit Cat state generation circuit generated by our approach is shown in Figure 16.

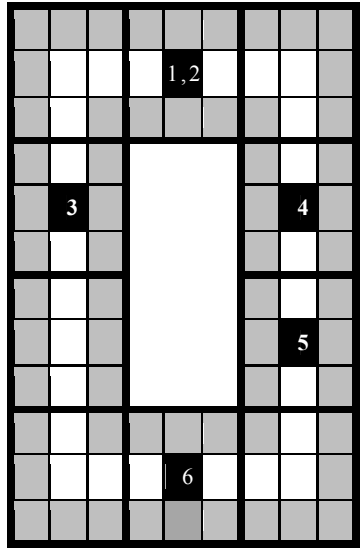

**Fig. 16** Layout and gate locations of 4-qubit Cat state generation circuit created by our approach

The initial placement of the qubits is also shown in Figure 17.

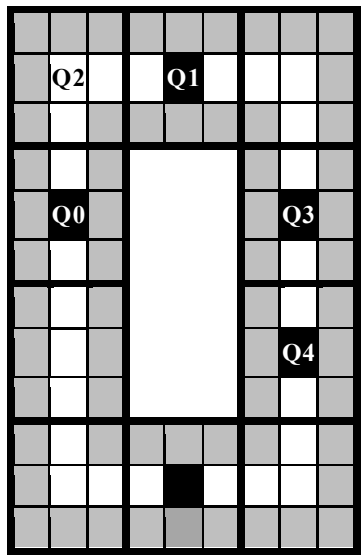

**Fig. 17** Initial placement of qubits for 4-qubit Cat state generation circuit

Since all the Cat state generation circuits have the same pattern, Table 12 can represent the general scheduling pattern of these circuits, the general pattern of the layout and the latency of these circuits as a function of number of qubits. As can be seen, the scheduling and layout pattern of these circuits have some differences for even and odd number of qubits.

**Table 12** Summary of scheduling, layout pattern and latency of Cat state generation circuit

| | n- qubit Cat state generation circuit n=2k+1 | n-qubit Cat state generation circuit n=2k |
|---|---|---|
| Number of instructions | n+2 | n+2 |
| Scheduling pattern | <table><tr><th>Stages of scheduling</th><th>Instructions which should be executed in this stage</th></tr><tr><td>1</td><td>1</td></tr><tr><td>2</td><td>2</td></tr><tr><td>3</td><td>3, 4</td></tr><tr><td>…</td><td>…</td></tr><tr><td>m</td><td>2m-3, 2m-2</td></tr><tr><td>…</td><td>…</td></tr><tr><td>(n+3)/2</td><td>n, n+1</td></tr><tr><td>(n+5)/2</td><td>n+2</td></tr></table> | <table><tr><th>Stages of scheduling</th><th>Instructions which should be executed in this stage</th></tr><tr><td>1</td><td>1</td></tr><tr><td>2</td><td>2</td></tr><tr><td>3</td><td>3, 4</td></tr><tr><td>…</td><td>…</td></tr><tr><td>m</td><td>2m-3, 2m-2</td></tr><tr><td>…</td><td>…</td></tr><tr><td>(n+4)/2</td><td>n+1</td></tr><tr><td>(n+6)/2</td><td>n+2</td></tr></table> |
| Layout graph | (layout graph with nodes 1,2 at top, branching to 3 and 4, continuing to n and n+1, joined at n+2; bracket labeled (n-1)/2) | (layout graph with nodes 1,2 at top, branching to 3 and 4, continuing to n-1 and n, then n+1, joined at n+2; bracket labeled (n-2)/2) |
| Parametric latency | $1 * t_H + 2 * t_{turn}$ $+ \left(\frac{n-1}{2} + 2\right) * t_{CNOT}$ $+ \left(\left(\frac{n-1}{2} + 4\right) * 3\right) * t_{straight\ move}$ | $1 * t_H + 2 * t_{turn}$ $+ \left(\frac{n}{2} + 2\right) * t_{CNOT}$ $+ \left(\left(\frac{n}{2} + 4\right) * 3\right) * t_{straight\ move}$ |

# 6- Conclusion

In this paper, we introduced a physical design flow consisting of a scheduler and a layout generator. The scheduler uses ILP and the layout generator uses a graph algorithm for generating a layout for quantum circuits with the aim of minimum latency. The experimental results show an improvement of about 11% in the latency for a set of the attempted benchmarks and about 9% for another set of benchmarks compared with the best in the literature. The approach presented in this paper can be used as basic methods of scheduling and layout generation for large circuits. Research on improving these methods for large circuits is underway. We also represented the general scheduling pattern of Cat state generation circuits, the general pattern of the layout and the latency of these circuits as a function of number of qubits.

**Acknowledgment**

We would like to thank Mehdi Saeedi from University of Southern California for his helpful discussion.